\pdfoutput=1
\documentclass[conference]{IEEEtran}
\IEEEoverridecommandlockouts
\usepackage{cite}
\usepackage{amsmath,amssymb,amsfonts}
\usepackage{graphicx}
\usepackage{textcomp}
\usepackage[table]{xcolor}
\usepackage{booktabs}
\usepackage{multirow}
\usepackage{siunitx}
\usepackage[T1]{fontenc}
\usepackage{array}
\usepackage{url}
\usepackage{tikz}
\usetikzlibrary{positioning,arrows.meta,fit,decorations.pathreplacing,shapes.geometric,backgrounds}
\newcommand{\gmid}{$g_m/I_D$}
\begin{document}

\title{CRAFT: Analog Circuit Design by Reasoning\\LLM Agents from Pre-Computed Lookup Tables}

\author{\IEEEauthorblockN{Hyunjun Park\textsuperscript{1,*}, Seungyub Han\textsuperscript{2,*}, Jason Jaesung Lee\textsuperscript{2}, Kahyun Kim\textsuperscript{2},\\
Soohyun Kim\textsuperscript{1}, Inwoo Nam\textsuperscript{1}, Sunmin Kim\textsuperscript{1}, Wooyoung Choi\textsuperscript{1} and Junhyung Kim\textsuperscript{2}}
\IEEEauthorblockA{\textsuperscript{1}Samsung Electronics, Republic of Korea\qquad \textsuperscript{2}Seoul National University, Seoul, Republic of Korea}
\thanks{\textsuperscript{*}These authors contributed equally to this work.}
}

\IEEEaftertitletext{\vspace{-12pt}}
\maketitle

\begin{abstract}
Automated analog circuit sizing methods, from black-box optimizers to recent large language model (LLM) agents,
typically search the design space with a SPICE simulator in the loop, whereas an experienced designer works out a design by hand
analysis. This paper presents CRAFT, a flow in which an LLM agent designs a given topology by calculation, as a designer does,
and without an initial sizing. The agent works only through DeX, a service that uses pre-computed lookup tables to provide device
characteristics and to solve the whole netlist for its operating point and small-signal response. It thus designs
without SPICE, and every final design is verified independently in HSPICE. We test CRAFT on four circuits in a
commercial \SI{28}{nm} process; the largest is a 37-transistor three-stage amplifier with 53 design variables, 20 specifications
and 74 saturation constraints. With Claude Opus~5.5, CRAFT meets every specification and constraint in all 20 runs, five per
circuit. On that amplifier, with DeX in place of HSPICE and an identical request for both, the same agent passes in 10 of 10 runs instead of 6 of 10 within an
hour and is 2.0$\times$ faster at the median.
\end{abstract}


\section{Introduction}
Generative AI, now driving innovation across the sciences, runs on silicon, and that silicon needs analog circuits:
the data converters, serial links and power regulators around its digital cores \cite{horowitz2014energy}.
Yet analog circuits are hard to design, because their performance metrics depend on many coupled parameters
and exhibit trade-offs. Analog design has therefore relied on designer expertise, slow to acquire and hard
to formalize. This paper asks how much of that expertise a language model can carry when it is
given the designer's own instruments.

One systematic method of analog design is the \gmid{} methodology, in which the designer calculates with
pre-computed lookup tables \cite{silveira1996gmid,jespers2017gmid,youssef2020lut,murmann2026isscc}. In this methodology, each transistor is assigned
a \gmid{} value according to its role in the circuit (differential pair, current mirror, cascode or output device). The other quantities the design needs are
then read from the tables as the calculation proceeds, which removes much of the inaccuracy of a hand calculation. The lookups themselves are
simple, but the design as a whole still demands expertise, because the designer must understand the circuit's
operation in detail and work through a web of parameterized equations and inequalities by a long chain of reasoning.
The \gmid{} coordinates give that reasoning a sound frame and ease the design somewhat, but the designer must still devise
the design procedure.

Another method treats sizing as an optimization problem. Optimization algorithms of many kinds take the design variables as
input and the simulated performance as output, learn the mapping between them, and search the design space with it
\cite{lyu2018bo,settaluri2020autockt,wang2020gcnrl,analoggymopt2026}. This approach needs no expert in
the loop, works for any topology and scales with compute; once the mapping is learned, it accounts for
every effect the device model captures. In the black-box setting considered here, a loss of saturation or cascode headroom appears as a violated metric, without an explicit diagnosis of its circuit-level cause. Such an algorithm's sample count grows steeply with the number of design variables, reaching thousands of simulations per design
\cite{analoggymopt2026}, and more for larger circuits.

\nocite{razavi2026pyramid,eesizer2025,agenticsizing2026,anaflow2025,whitebox2025,bujana2026selfcal,adollm2024,autosizer2026,xing2025astra,panda2025,artisan2024,whiteop2026}%
\begin{figure}[t]
\centering
\includegraphics[width=\columnwidth]{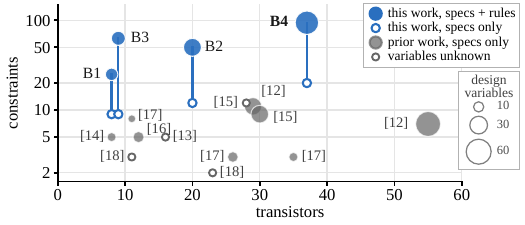}\vspace{-2pt}
\caption{The largest circuits in seven prior works on LLM sizing \cite{agenticsizing2026,bujana2026selfcal,adollm2024,autosizer2026,xing2025astra,anaflow2025,whitebox2025}
and Benchmarks 1--4 (B1--B4). B1--B4 also count device rules as constraints.}
\label{fig:size}
\vspace{-6pt}
\end{figure}

\begin{figure*}[t]
\centering
\def\HB{4.6mm}\def\P{2mm}%
\newcolumntype{R}{!{\color{black!35}\vrule width 0.4pt\hspace{3pt}}}%
\newcolumntype{B}{>{\bfseries\boldmath}l}%
\ifdefined\dexA\else\newsavebox\dexA\newsavebox\dexB\newsavebox\dexC\newlength\dexG\newlength\dexH\fi
\sbox\dexA{\footnotesize\renewcommand{\arraystretch}{1.04}\begin{tabular}{@{}B@{\hspace{3pt}}R@{}l@{}}
  grid    & $\mathbf x=(L,W,V_{GS},V_{DS},V_{BS})$\\
  stored  & $q(\mathbf x)$ for $q\in\{I_D,\,g_m,\,g_{ds},\,g_{mb},$\\
          & \quad $V_{th},\,V_{DSAT},\,\partial Q_i/\partial V_j,$\\
          & \quad $g_m/I_D,\,J_D,\,g_m/g_{ds},\,f_T\}$
  \end{tabular}}%
\sbox\dexB{\footnotesize\renewcommand{\arraystretch}{1.04}\begin{tabular}{@{}B@{\hspace{3pt}}R@{}l@{}}
  lookup           & $q(\mathbf x)$, interpolated\\
  inverse $V_{GS}$ & $V_{GS}=q^{-1}(q^{*})$\\
  cross lookup     & $q_2$ at $V_{GS}=q_1^{-1}(q_1^{*})$\\
  width sizing     & $W=I_D^{-1}(I_D^{*})$
  \end{tabular}}%
\sbox\dexC{\footnotesize\renewcommand{\arraystretch}{1.04}\begin{tabular}{@{}B@{\hspace{3pt}}R@{}l@{}}
  DC    & KCL $\mathbf r(\mathbf v)=\mathbf 0$, $\Delta\mathbf v=-\mathbf G(\mathbf v)^{-1}\mathbf r(\mathbf v)$\\
  AC    & $(\mathbf G+s\mathbf C)\,\mathbf v=\mathbf b$ $\rightarrow$ $H(s)$, loop gain, PM\\
  poles & $\det(\mathbf G+s\mathbf C)=0$\\
  zeros & $N(s)=0$, $H(s)=N(s)/\det(\mathbf G+s\mathbf C)$
  \end{tabular}}%
\setlength\dexH{\ht\dexA}\addtolength\dexH{\dp\dexA}%
\ifdim\dimexpr\ht\dexB+\dp\dexB\relax>\dexH \setlength\dexH{\dimexpr\ht\dexB+\dp\dexB\relax}\fi
\ifdim\dimexpr\ht\dexC+\dp\dexC\relax>\dexH \setlength\dexH{\dimexpr\ht\dexC+\dp\dexC\relax}\fi
\setlength\dexH{\dimexpr\dexH+\HB+1.4mm\relax}\def\H{\dexH}%
\setlength\dexG{\dimexpr(\textwidth-1pt-\wd\dexA-\wd\dexB-\wd\dexC-\P*6)/2\relax}%
\begin{tikzpicture}[x=1cm, y=1cm, font=\footnotesize,
  arr/.style={-{Latex[length=1.8mm]}, line width=0.7pt, black!70}]
\def\panel#1#2#3#4#5#6{%
  \path (#2) coordinate (#1-nw) ++(\wd#3+2*\P,-\H) coordinate (#1-se);
  \fill[#4, rounded corners=2pt] (#1-nw) rectangle (#1-se);
  \begin{scope}\clip[rounded corners=2pt] (#1-nw) rectangle (#1-se);
    \fill[#5] (#1-nw) rectangle ([yshift=-\HB]#1-nw -| #1-se);\end{scope}
  \draw[black!35, line width=0.4pt] ([yshift=-\HB]#1-nw) -- ([yshift=-\HB]#1-nw -| #1-se);
  \draw[black!60, line width=0.5pt, rounded corners=2pt] (#1-nw) rectangle (#1-se);
  \node[inner sep=0pt, fit=(#1-nw)(#1-se)] (#1) {};
  \node[anchor=west, inner sep=0pt] at ([xshift=\P,yshift=-\HB/2]#1-nw) {\small\strut\textbf{#6}};
  \node[anchor=west, inner sep=0pt] at ([xshift=\P,yshift=-\HB/2-\H/2]#1-nw) {\usebox#3};}
\panel{tab}{0,0}{\dexA}{black!3}{black!10}{5-D table}
\panel{dq}{[xshift=\dexG]tab-nw -| tab-se}{\dexB}{blue!4}{blue!14}{device queries}
\panel{cs}{[xshift=\dexG]dq-nw -| dq-se}{\dexC}{orange!5}{orange!20}{circuit solver}
\draw[arr] ([xshift=0.8pt]tab.east) -- ([xshift=-0.8pt]dq.west);
\draw[arr] ([xshift=0.8pt]dq.east) -- ([xshift=-0.8pt]cs.west);
\node[inner sep=0pt, fit=(tab)(dq)(cs)] (row) {};
\node[draw=black!60, rounded corners=2pt, align=left, inner sep=2.5pt, fill=white, line width=0.5pt, anchor=north west, text width=\dimexpr\textwidth-10pt\relax] (man) at ([yshift=-6.0mm]row.south west) {\textbf{\small manual}\\[2pt]
  {\ttfamily\small dex-query lookup\ \ device=nch quantity=gm\_id L=\textrm{\dots} vgs=\textrm{\dots} vds=\textrm{\dots}\ \ $\rightarrow$\ \ \{"values": [\textrm{\dots}]\}\\
  dex-query dc\ \ \ \ \ \ -{}-file circuit.json\ \ $\rightarrow$\ \ \{"nodes": \{\textrm{\dots}\}, "devices": \{\textrm{\dots}\}\}\\
  dex-query ac\,|\,pz\ \ -{}-file circuit.json\ \ $\rightarrow$\ \ \{"gain": [\textrm{\dots}], "poles": [\textrm{\dots}]\}}};
\draw[decorate, decoration={brace, mirror, amplitude=1.2mm}, black!60, line width=0.5pt] ([yshift=-0.6mm]row.south west) -- ([yshift=-0.6mm]row.south east);
\draw[arr, {Latex[length=1.8mm]}-{Latex[length=1.8mm]}] (man.north -| row.center) -- ([yshift=-2.0mm]row.south -| row.center)
  node[midway, right=3pt, font=\footnotesize\itshape, text=black!70] {the agent reaches all of the above only through the manual's commands};
\end{tikzpicture}%
\caption{DeX: the tables, the device queries and the circuit solver, reached only through the generated manual.}
\label{fig:dex}
\end{figure*}
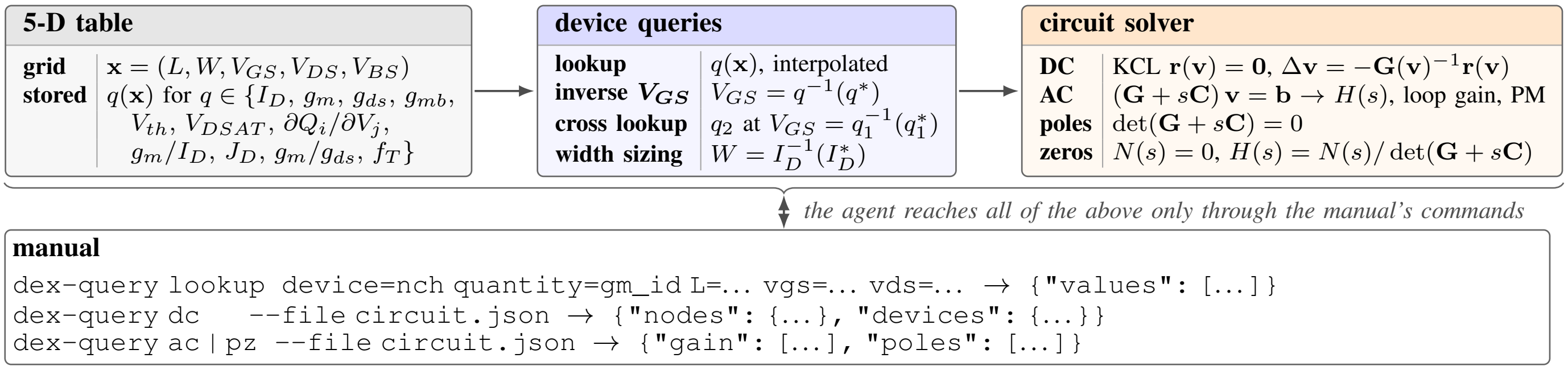

Both methods aim at the same goal: designing a given topology, the third of the five levels of analog design
\cite{razavi2026pyramid}. That level rests on the first two: analyzing the circuit
correctly (level~1) and drawing on intuition about it (level~2). The lookup-table method reaches that level by climbing the first three levels
in order, whereas the optimizer goes straight to the third level. The two levels the
optimizer skips are exactly where language models are already being tested \cite{razavi2026pyramid}. This
paper proposes a way for a large language model (LLM) agent to climb from the second level to the third, through
a service we call DeX, the Device eXplorer. DeX holds pre-computed lookup tables and a solver that uses them to analyze
the whole netlist, so that the agent can plan the design procedure itself. We call the method CRAFT, Circuit design by Reasoning LLM Agents From pre-computed lookup Tables. The agent
receives a request and decides what to calculate, what to check and when the design is done. Whether the
final design meets the specification is then judged outside the agent, in HSPICE.

Recent LLM agents for sizing share this goal and differ in how they make the climb. In one line of work, the
agent predicts the next design change from circuit and device knowledge \cite{eesizer2025,agenticsizing2026,anaflow2025,whitebox2025} or from
equations it writes and calibrates \cite{bujana2026selfcal}, makes the change, and corrects the prediction by
simulation before the next step. In a second line of work, the agent steers a numerical optimizer, and the optimizer makes the climb by simulation.
The agent seeds the optimizer \cite{adollm2024}, chooses the variables and algorithm \cite{autosizer2026}, fixes the \gmid{} of
each transistor \cite{xing2025astra}, writes the constraints \cite{panda2025} or calls the optimizer when its own reasoning
stalls \cite{anaflow2025}. In a third line of work, the agent designs a model comprising a few behavioral stages with simulation in the loop. The result is mapped to transistors only at the end
\cite{artisan2024,whiteop2026}. In CRAFT, the agent uses device queries and circuit analyses based on pre-computed tables to size circuits under multiple simultaneous constraints. Fig.~\ref{fig:size} compares our benchmarks with the largest circuits sized in seven prior studies of LLM agents.
Our largest circuit, Benchmark~4, has 53 design variables, close to the 60 variables of the largest circuit in \cite{agenticsizing2026}, yet must satisfy 20 specifications and 74 device rules, against 7 specifications for that circuit and at most 12 for any circuit in \cite{agenticsizing2026,bujana2026selfcal,adollm2024,autosizer2026,xing2025astra,anaflow2025,whitebox2025}.

This paper makes three contributions.
\begin{itemize}\setlength{\itemsep}{0pt}\setlength{\topsep}{0pt}
\item We introduce DeX, a service that combines device queries and inverse sizing with whole-netlist DC and small-signal analysis using pre-computed lookup tables. DeX connects device-level design calculations to circuit-level evaluation through a unified interface, replacing SPICE simulations in the sizing loop for the supported analyses.
\item We develop CRAFT, an autonomous sizing flow in which an LLM plans its own design procedure from a given topology and specifications, without a designer-provided initial sizing. The agent plans node voltages and branch currents, obtains device sizes and analysis results through DeX, and uses the results to diagnose and resolve violations of performance specifications and device rules.
\item We evaluate CRAFT on four circuits in a commercial 28-nm process, independently checking every final design in HSPICE. With Opus~5.5, CRAFT satisfies every specification and device rule in all 20 runs. In a separate comparison on the largest, 53-variable circuit using the same agent and request, the agent succeeds in 10/10 runs with DeX versus 6/10 with HSPICE within a 60-minute budget, reducing median time to a passing design from 57.7 to 28.2 minutes. A hybrid DeX--ngspice configuration also achieves 100\,\% success on the Hard split of AMS-SizingBench \cite{autosizer2026} at approximately one-quarter of AutoSizer's mean runtime with the same LLM.
\end{itemize}%

\section{DeX: the Device eXplorer service}\label{sec:dex}
The lookup-table method of sizing \cite{silveira1996gmid,jespers2017gmid,murmann2026isscc} rests on
quantities that keep their physical meaning under any device model. These quantities are tabulated once per process and consulted at every
step of the design. DeX answers an agent's device queries from these tables through a text interface, and adds a
simple Newton solver that evaluates the whole netlist with the same tables, in place of the hand calculation of each spec
\cite{youssef2020lut,murmann2024script} (Fig.~\ref{fig:dex}).

\begin{figure}[t]
\centering
\sbox0{\includegraphics{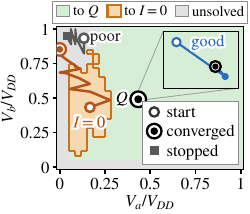}}%
\begin{minipage}[b]{0.41\columnwidth}\centering\parbox[b][\ht0][c]{\linewidth}{\centering\includegraphics[height=98pt]{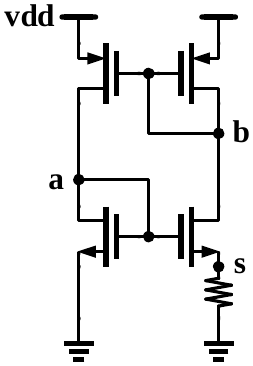}}\\[3pt]
\footnotesize (a) $\beta$-multiplier reference\end{minipage}\hfill
\begin{minipage}[b]{0.58\columnwidth}\centering\usebox0\\[3pt]
\footnotesize (b) Newton--Raphson from three starts\end{minipage}
\caption{A toy example of the Newton--Raphson method: (a) a reference circuit with two solutions, $Q$ and $I=0$,
and (b) its $V_a$--$V_b$ plane, shaded by the outcome from each start ($V_s=\SI{50}{mV}$), with the paths
from three starts.}
\label{fig:newton}
\end{figure}

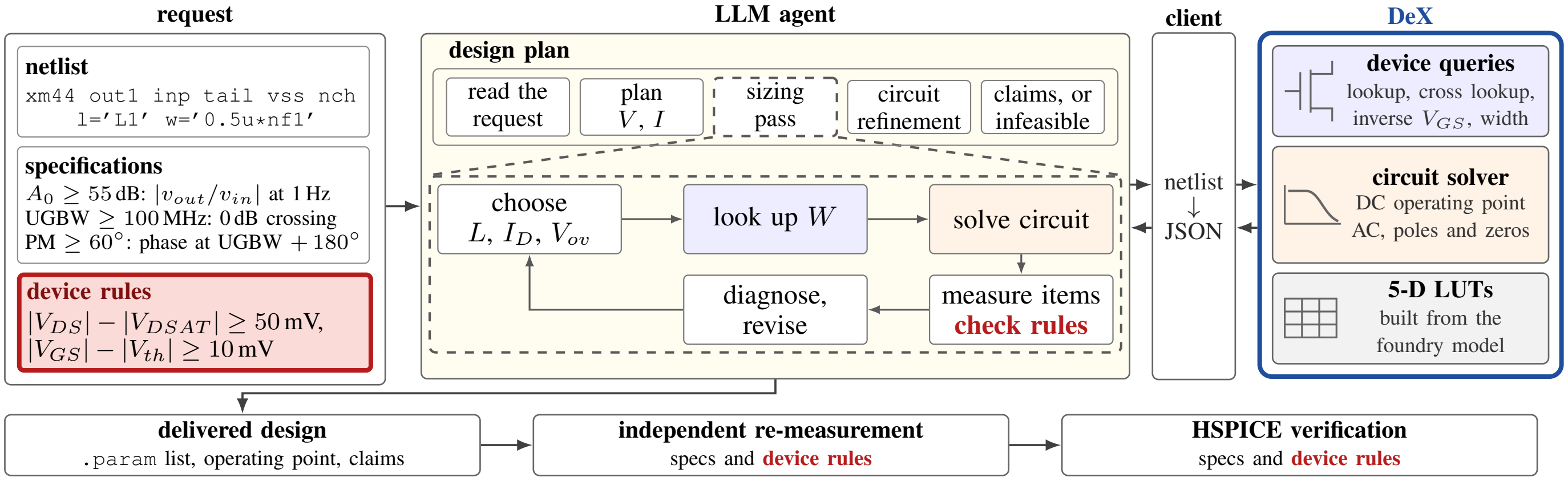
\begin{figure*}[t]
\centering
\definecolor{dexedge}{HTML}{1F4E9C}%
\definecolor{rulefill}{HTML}{FDE7CF}\definecolor{ruleedge}{HTML}{B71C1C}\definecolor{rulecard}{HTML}{FADBD8}\definecolor{ruledark}{HTML}{7F1010}%
\begin{tikzpicture}[x=1cm, y=1cm, font=\footnotesize,
  box/.style={draw=black!60, rounded corners=2pt, align=center, inner sep=3pt, fill=white, line width=0.5pt, text width=#1},
  card/.style={draw=black!45, rounded corners=1.5pt, align=flush left, inner sep=2.5pt, fill=white, line width=0.4pt, font=\scriptsize, text width=39mm},
  stp/.style={draw=black!55, rounded corners=1.5pt, align=center, inner sep=2.5pt, fill=white, line width=0.4pt, font=\small, text width=19.5mm, minimum height=8mm},
  chip/.style={draw=black!50, rounded corners=1.5pt, align=center, inner sep=1.5pt, fill=white, line width=0.4pt, font=\footnotesize, text width=13.2mm},
  zone/.style={draw=black!50, rounded corners=4pt, inner sep=5pt, line width=0.6pt},
  ttl/.style={anchor=south west, font=\footnotesize\bfseries, inner sep=1pt},
  note/.style={font=\scriptsize\itshape, text=black!75, align=center, inner sep=1pt},
  num/.style={circle, draw=red!70!black, text=red!70!black, font=\scriptsize\bfseries, inner sep=0.5pt, minimum size=3.6mm, fill=white},
  arr/.style={-{Latex[length=1.8mm]}, line width=0.7pt, black!75},
  darr/.style={{Latex[length=1.8mm]}-{Latex[length=1.8mm]}, line width=0.7pt, black!75},
  larr/.style={-{Latex[length=1.4mm]}, line width=0.6pt, black!65},
  ico/.style={line width=0.6pt, black!60},
  pics/mosfet/.style={code={\draw[ico] (0,0.25) -- (0,-0.25) (-0.12,0.22) -- (-0.12,-0.22) (-0.3,0) -- (-0.12,0)
      (0,0.2) -- (0.25,0.2) -- (0.25,0.35) (0,-0.2) -- (0.25,-0.2) -- (0.25,-0.35);}},
  pics/grid/.style={code={\draw[ico] (0,0) rectangle (0.6,0.45); \draw[ico] (0.2,0) -- (0.2,0.45) (0.4,0) -- (0.4,0.45) (0,0.15) -- (0.6,0.15) (0,0.3) -- (0.6,0.3);}},
  pics/bode/.style={code={\draw[ico] (0,0) -- (0,0.45) (0,0) -- (0.65,0); \draw[ico, line width=0.8pt] (0.04,0.38) -- (0.25,0.38) .. controls (0.4,0.38) and (0.45,0.1) .. (0.62,0.03);}}]
\node[card, anchor=north west, minimum height=10.5mm] (c1) at (0,1.494) {{\footnotesize\bfseries netlist}\\[1pt]
  {\fontsize{7}{8}\selectfont\ttfamily xm44 out1 inp tail vss nch\\\ \ \ \ l='L1' w='0.5u*nf1'}};
\node[card, anchor=north west, minimum height=13.5mm] (c2) at ([yshift=-1mm]c1.south west) {{\footnotesize\bfseries specifications}\\[1pt]
  $A_0 \ge 55$\,dB: $|v_{out}/v_{in}|$ at 1\,Hz\\
  UGBW $\ge 100$\,MHz: 0\,dB crossing\\
  PM $\ge 60^\circ$: phase at UGBW $+\,180^\circ$};
\node[card, anchor=north west, minimum height=10.5mm, draw=ruleedge, fill=rulecard, line width=1.5pt, font=\footnotesize] (c3) at ([yshift=-1mm]c2.south west) {{\bfseries\color{ruledark} device rules}\\[1pt]
  $|V_{DS}|-|V_{DSAT}|\ge 50$\,mV,\ $|V_{GS}|-|V_{th}|\ge 10$\,mV};
\node[box=0mm, draw=black!60, fit=(c1)(c2)(c3), inner sep=4pt, fill=none] (req) {};
\node[ttl, anchor=south] at ([yshift=0.3mm]req.north) {request};
\node[box=80mm, anchor=north west, fill=yellow!7, minimum height=40mm] (agent) at ([xshift=4mm]req.north east) {};
\node[ttl, anchor=south] at ([yshift=0.3mm]agent.north) {LLM agent};
\node[chip] (a1) at ([xshift=-31mm, yshift=11.5mm]agent.center) {read the\\request};
\node[chip] (a2) at ([xshift=-15.5mm, yshift=11.5mm]agent.center) {plan\\$V$, $I$};
\node[chip, draw=black!65, dashed, line width=0.8pt] (a3) at ([xshift=0mm, yshift=11.5mm]agent.center) {sizing\\pass};
\node[chip] (a4) at ([xshift=15.5mm, yshift=11.5mm]agent.center) {circuit\\refinement};
\node[chip] (a5) at ([xshift=31mm, yshift=11.5mm]agent.center) {claims, or\\infeasible};
\node[stp] (s1) at ([xshift=-28.5mm, yshift=-1.5mm]agent.center) {choose\\$L$, $I_D$, $V_{ov}$};
\node[stp, fill=blue!7] (s2) at ([xshift=0mm, yshift=-1.5mm]agent.center) {look up $W$};
\node[stp, fill=orange!10] (s3) at ([xshift=28.5mm, yshift=-1.5mm]agent.center) {solve circuit};
\node[stp] (s4) at ([xshift=28.5mm, yshift=-12mm]agent.center) {measure items\\{\bfseries\color{ruleedge}check rules}};
\node[stp] (s5) at ([xshift=0mm, yshift=-12mm]agent.center) {diagnose,\\revise};
\draw[larr] (s1) -- (s2);
\draw[larr] (s2) -- (s3);
\draw[larr] (s3) -- (s4);
\draw[larr] (s4) -- (s5);
\draw[larr] (s5.west) -| (s1.south);
\node[draw=black!65, dashed, line width=0.8pt, rounded corners=2pt, inner sep=2.5pt, fit=(s1)(s3)(s4)(s5)] (zoom) {};
\draw[black!65, dashed, line width=0.8pt] (a3.south west) -- (zoom.north west);
\draw[black!65, dashed, line width=0.8pt] (a3.south east) -- (zoom.north east);
\node[draw=black!50, rounded corners=2pt, inner xsep=1.5mm, inner ysep=1mm, fit=(a1)(a5)] (plan) {};
\node[ttl, font=\footnotesize\bfseries] at ([xshift=1.5mm, yshift=0.2mm]plan.north west) {design plan};
\draw[arr] (req.east |- agent.center) -- (agent.west |- agent.center);
\node[box=0mm, minimum width=32mm, minimum height=10.5mm, anchor=north west, fill=blue!7] (dq) at ([xshift=16.4mm, yshift=-1.41mm]agent.north east) {};
\node[align=center, text width=27mm] at ([xshift=3.5mm]dq.center) {\textbf{device queries}\\{\scriptsize\color{black!80} lookup, cross lookup, inverse $V_{GS}$, width}};
\node[box=0mm, minimum width=32mm, minimum height=13.5mm, anchor=north west, fill=orange!10] (cs) at ([yshift=-1mm]dq.south west) {};
\node[align=center, text width=27mm] at ([xshift=3.5mm]cs.center) {\textbf{circuit solver}\\{\scriptsize\color{black!80} DC operating point\\AC, poles and zeros}};
\node[box=0mm, minimum width=32mm, minimum height=10.5mm, anchor=north west, fill=black!5] (lut) at ([yshift=-1mm]cs.south west) {};
\node[align=center, text width=27mm] at ([xshift=3.5mm]lut.center) {\textbf{5-D LUTs}\\{\scriptsize\color{black!80} built from the foundry model}};
\pic at ([xshift=4.5mm]dq.west) {mosfet};
\pic at ([xshift=1.5mm, yshift=-2.2mm]cs.west) {bode};
\pic at ([xshift=1.5mm, yshift=-2.2mm]lut.west) {grid};
\node[zone, inner sep=4pt, fit=(dq)(cs)(lut), draw=dexedge, line width=1.5pt] (dex) {};
\node[ttl, anchor=south, text=dexedge] at ([yshift=0.3mm]dex.north) {DeX};
\path (agent.east |- agent.center) -- coordinate[pos=0.5] (cmid) (dex.west |- agent.center);
\path (cmid |- agent.north) coordinate (ctop) (cmid |- agent.south) coordinate (cbot);
\node[draw=black!60, rounded corners=2pt, fill=white, line width=0.5pt, inner sep=0pt, minimum width=9.6mm, fit=(ctop)(cbot)] (cli) {};
\node[ttl, anchor=south] at ([yshift=0.3mm]cli.north) {client};
\node[align=center, inner sep=0pt, font=\footnotesize, text=black!85] at (cli.center) {netlist\\[-1pt]$\downarrow$\\[-0.5pt]JSON};
\draw[arr] ([yshift=2.5mm]agent.east |- agent.center) -- ([yshift=2.5mm]cli.west |- agent.center);
\draw[arr] ([yshift=2.5mm]cli.east |- agent.center) -- ([yshift=2.5mm]dex.west |- agent.center);
\draw[arr] ([yshift=-2.5mm]dex.west |- agent.center) -- ([yshift=-2.5mm]cli.east |- agent.center);
\draw[arr] ([yshift=-2.5mm]cli.west |- agent.center) -- ([yshift=-2.5mm]agent.east |- agent.center);
\path (req.west); \pgfgetlastxy{\rx}{\ry}
\path (dex.east); \pgfgetlastxy{\dx}{\dy}
\pgfmathsetlengthmacro{\bw}{(\dx-\rx-12mm-4pt)/3}
\node[box=0mm, text width=\dimexpr\bw-6pt\relax, anchor=north west, inner sep=2pt] (out) at ([yshift=-4mm]req.west |- agent.south) {\textbf{delivered design}\\{\scriptsize \texttt{.param} list, operating point, claims}};
\node[box=0mm, text width=\dimexpr\bw-6pt\relax, anchor=north west, inner sep=2pt] (ev) at ([xshift=6mm]out.north east) {\textbf{independent re-measurement}\\{\scriptsize specs and {\bfseries\color{ruleedge}device rules}}};
\node[box=0mm, text width=\dimexpr\bw-6pt\relax, anchor=north west, inner sep=2pt] (hs) at ([xshift=6mm]ev.north east) {\textbf{HSPICE verification}\\{\scriptsize specs and {\bfseries\color{ruleedge}device rules}}};
\draw[arr] (agent.south) -- ++(0,-1.6mm) -| (out.north);
\draw[arr] (out) -- (ev);
\draw[arr] (ev) -- (hs);
\end{tikzpicture}%
\caption{Overview of CRAFT with DeX. The agent reads the request, plans the design and obtains the information it needs from DeX through the client. The delivered design is then re-measured independently of the agent and verified in HSPICE.}
\label{fig:flow}
\end{figure*}

\textbf{Tables.} Each device type is characterized once on the five-dimensional grid
$\mathbf x=(L,W,V_{GS},V_{DS},V_{BS})$ (Fig.~\ref{fig:dex}, left). At each point the table holds the drain current, the
small-signal parameters, the charge derivatives and pre-computed design quantities such as \gmid{}, $J_D=I_D/W$, $g_m/g_{ds}$ and
$f_T$, all read by piecewise-polynomial interpolation. Devices wider than the table's width range are split into parallel
units.

\textbf{Device queries.} Every stored or derived quantity is a function $q(\mathbf x)$ on the grid, and the four
queries are operations on such functions with all but one coordinate of $\mathbf x$ held fixed (Fig.~\ref{fig:dex},
middle). \emph{Lookup} evaluates $q(\mathbf x)$ at a point or on a grid of points. \emph{Inverse $V_{GS}$} returns
$V_{GS}=q^{-1}(q^{*})$ at fixed $(L,W,V_{DS},V_{BS})$, for a $q$ that is monotone in $V_{GS}$, such as $g_m/I_D$ or
$J_D$. \emph{Cross lookup} composes the two, returning a second quantity $q_2$ at $V_{GS}=q_1^{-1}(q_1^{*})$ for a target
$q_1^{*}$ of the first. \emph{Width sizing}
returns the width at which the device carries a target current, $W=I_D^{-1}(I_D^{*})$ along the width axis at fixed
$(L,V_{GS},V_{DS},V_{BS})$.

\textbf{Circuit solver.} The DC solve finds the node voltages $\mathbf v$ from Kirchhoff's current law $\mathbf r(\mathbf v)=\mathbf 0$ at
every unknown node by damped Newton--Raphson iteration, $\Delta\mathbf v=-\mathbf G(\mathbf v)^{-1}\mathbf r(\mathbf v)$,
with $\mathbf G$ assembled from the interpolated $g_m$, $g_{ds}$, $g_{mb}$ and the resistors (Fig.~\ref{fig:dex},
right). Each step is limited to \SI{0.1}{V} per node and halved until the residual falls, and a point is accepted when
every node's residual is within \SI{1}{nA} plus $10^{-6}$ of the current through the node. The solver does not use $g_{\min}$ or source stepping,
because the agent plans the node voltages and branch currents first and reads each transistor's width for its planned $L$, $I_D$, $V_{GS}$, $V_{DS}$ and $V_{BS}$. The planned voltages thus already satisfy Kirchhoff's current law up to small errors, and Newton--Raphson started from there can converge in a few steps to the intended operating point. For
example, the self-biased reference of Fig.~\ref{fig:newton}(a) has two solutions, the intended operating point $Q$ and a
zero-current state, and Fig.~\ref{fig:newton}(b) shows the outcome of Newton--Raphson alone from each start. From starts in the
orange region it settles in the zero-current state, the other solution of the circuit, and from starts in the unsolved
region it does not converge. From the planned start (good), close to $Q$, it converges in four steps. At the converged operating point the solver reports, for every transistor,
the bias, $V_{th}$, $V_{DSAT}$, the saturation margin $|V_{DS}|-|V_{DSAT}|$ and whether the point lies inside
the table. At that point it assembles the small-signal system $(\mathbf G+s\mathbf C)\,\mathbf v=\mathbf b$, with $\mathbf C$ the
charge-conserving capacitance matrix from the stored $\partial Q_i/\partial V_j$ and $\mathbf b$ the vector of currents injected
into the nodes when the sources are applied. It returns the transfer function
$H(s)$ from the sources to any node, the unity-gain frequency and phase margin read from $H(s)$, the poles from
$\det(\mathbf G+s\mathbf C)=0$ with the nodes associated with each pole, and the zeros from $N(s)=0$, with
$H(s)=N(s)/\det(\mathbf G+s\mathbf C)$. Table interpolation introduces some error, but
the solution stays far closer to the device model than a hand calculation \cite{jespers2017gmid,youssef2020lut,murmann2026isscc} and is accurate enough for design
(Section~\ref{sec:outcome}).

\textbf{Interface.} An agent reaches DeX only through a client that sends JSON queries and receives JSON results. A
converter in the client, run locally, reads the designer's SPICE testbench netlist together with a \texttt{.param} file
of sizes and writes the circuit in the solver's own format. The agent therefore never retypes the circuit, and the independent evaluator of Section~\ref{sec:craft} later re-measures
the same netlist through the same converter. The server validates each query against one file of query definitions before executing the query, and the
manual is generated from the same file
(Fig.~\ref{fig:dex}, bottom). The agent never sees the tables, only query results.

\section{CRAFT: the flow and the protocol}\label{sec:craft}
In CRAFT, an LLM agent uses DeX's device queries and circuit analyses to plan and refine circuit designs.
Fig.~\ref{fig:flow} summarizes the protocol in three parts: the design request, the agent's sizing process, and
independent evaluation.

\newlength{\schh}
\setlength{\schh}{\dimexpr(\textwidth-1.2cm)*1000/5107\relax}
\newcommand{\schematic}[3]{\begin{minipage}[b]{#2}\centering#1\\[1pt]\footnotesize #3\end{minipage}}
\begin{figure*}[t]
\centering
\schematic{\includegraphics[width=\linewidth]{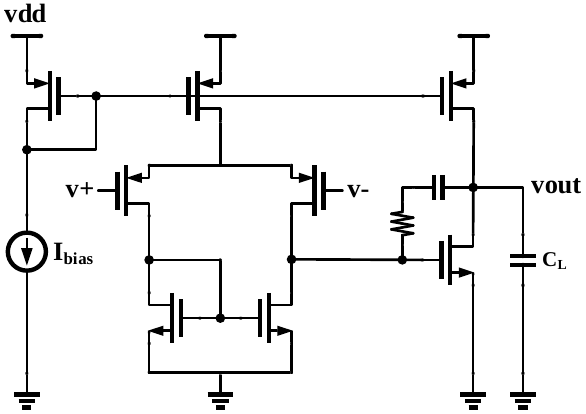}}{\dimexpr\schh*2809/1987\relax}{(a) Benchmark 1: two-stage op-amp}\hfill
\schematic{\includegraphics[width=\linewidth]{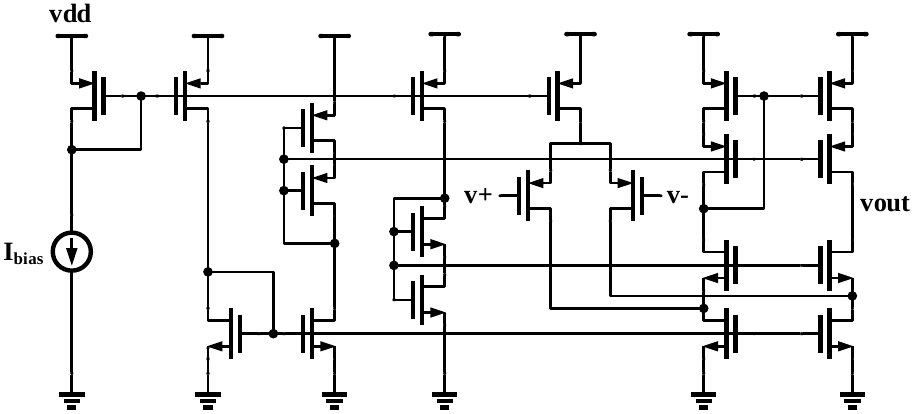}}{\dimexpr\schh*4388/1987\relax}{(b) Benchmark 2: folded-cascode amplifier}\hfill
\schematic{\includegraphics[width=\linewidth]{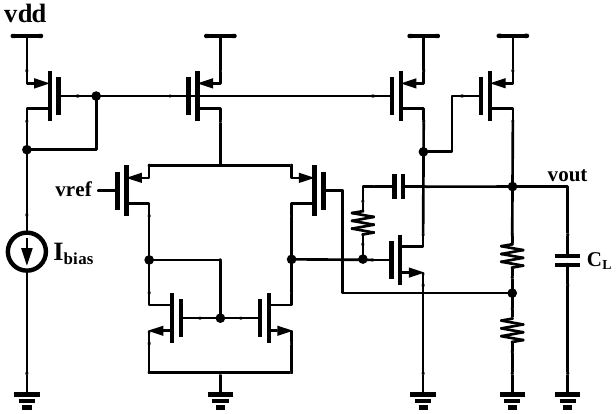}}{\dimexpr\schh*2952/1987\relax}{(c) Benchmark 3: LDO regulator}

\vspace{6pt}
\schematic{\includegraphics[width=\linewidth]{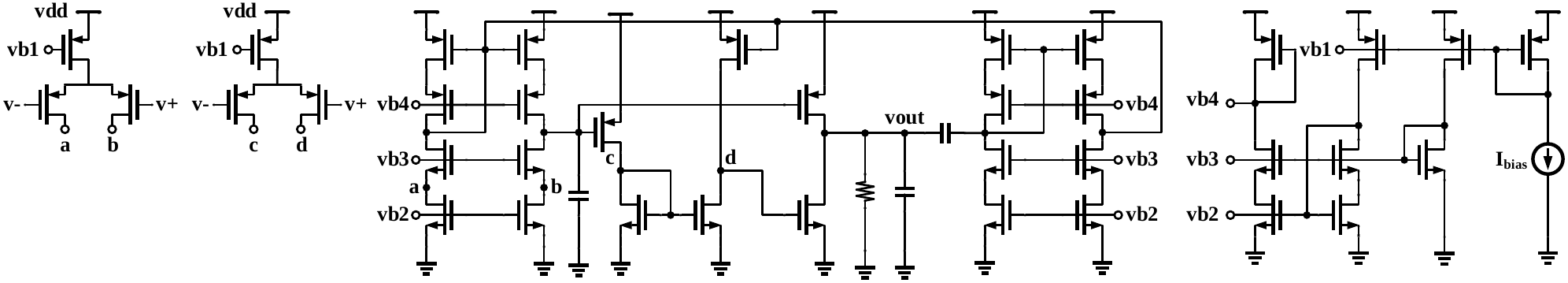}}{\textwidth}{(d) Benchmark 4: DACFC three-stage amplifier}%
\makebox[0pt][r]{\makebox[\textwidth][l]{\raisebox{1pt}[0pt][0pt]{\setlength{\fboxsep}{1.5pt}\setlength{\fboxrule}{0.7pt}%
\fcolorbox{black!45}{white}{\small\setlength{\tabcolsep}{1.8pt}\renewcommand{\arraystretch}{0.92}%
\begin{tabular}[b]{@{}lcccc@{}}
Benchmark & 1 & 2 & 3 & 4\\\specialrule{0.4pt}{1pt}{1.5pt}
Transistors & 8 & 20 & 9 & 37\\
Design variables & 15 & 31 & 19 & \textbf{53}\\
Specs & 9 & 12 & 9 & \textbf{20}\\
Device rules & 16 & 38 & 54 & \textbf{74}
\end{tabular}}}}}%
\caption{The four benchmark circuits: (a) $V_{CM}=\SI{0.62}{V}$, $C_L=\SI{1}{pF}$; (b) $V_{CM}=\SI{0.20}{V}$,
$C_L=\SI{1}{pF}$; (c) $V_{ref}=\SI{0.28}{V}$, $C_L=\SI{500}{pF}$, $I_L=0.5$/0.75/\SI{1}{mA}; (d) $V_{CM}=\SI{0.32}{V}$,
$C_L=60/100/160\,\si{pF}\parallel\SI{5}{k\ohm}$.}
\label{fig:schematics}
\end{figure*}
\begin{table*}[t]
\caption{Performance metrics for each circuit.}
\label{tab:specs}
\centering\fontsize{7.5}{8.6}\selectfont\setlength{\tabcolsep}{2pt}%
\renewcommand{\arraystretch}{1.0}
\begin{tabular*}{\textwidth}{@{\extracolsep{\fill}}lp{0.935\textwidth}@{}}
\toprule
Circuit & Metrics and limits \\
\midrule
B1 & \mbox{$I_Q$\,$\le$\,110\,\textmu A}\hspace{0.3em}\mbox{$A_0$\,$\ge$\,55\,dB}\hspace{0.3em}\mbox{CMRR\,$\ge$\,50\,dB}\hspace{0.3em}\mbox{UGBW\,$\ge$\,100\,MHz}\hspace{0.3em}\mbox{PM\,$\ge$\,60$^\circ$}\hspace{0.3em}\mbox{SR\,$\ge$\,20\,V/\textmu s}\hspace{0.3em}\mbox{$|V_{os}|$\,$\le$\,1\,mV}\hspace{0.3em}\mbox{Swing\,$\ge$\,0.5\,V$_{pp}$} \\
B2 & \mbox{$I_Q$\,$\le$\,400\,\textmu A}\hspace{0.3em}\mbox{$A_0$\,$\ge$\,54\,dB}\hspace{0.3em}\mbox{CMRR\,$\ge$\,60\,dB}\hspace{0.3em}\mbox{PSRR\,$\ge$\,48\,dB}\hspace{0.3em}\mbox{GBW\,$\ge$\,220\,MHz}\hspace{0.3em}\mbox{PM\,$\ge$\,70$^\circ$}\hspace{0.3em}\mbox{SR\,$\ge$\,90\,V/\textmu s}\hspace{0.3em}\mbox{$T_s$\,$\le$\,3.7\,ns}\hspace{0.3em}\mbox{$T_r$\,$\le$\,1.7\,ns}\hspace{0.3em}\mbox{$|V_{os}|$\,$\le$\,1\,mV}\hspace{0.3em}\mbox{Swing\,$\ge$\,0.2\,V$_{pp}$} \\
B3 & \mbox{$I_Q$\,$\le$\,70\,\textmu A}\hspace{0.3em}\mbox{PM\,$\ge$\,60$^\circ$}\hspace{0.3em}\mbox{Efficiency\,$\ge$\,67.5\,\%}\hspace{0.3em}\mbox{Dropout\,$\le$\,80\,mV}\hspace{0.3em}\mbox{Load reg.\,$\le$\,0.4\,\%}\hspace{0.3em}\mbox{Line reg.\,$\le$\,0.25\,\%}\hspace{0.3em}\mbox{PSRR\,$\ge$\,30\,dB}\hspace{0.3em}\mbox{$|\Delta V_{out}|$\,$\le$\,7\,mV} \\
B4 & \mbox{$A_0$\,$\ge$\,60\,dB}\hspace{0.3em}\mbox{GBW\,$\ge$\,80\,MHz}\hspace{0.3em}{\mbox{PM 60 pF\,$\ge$\,50$^\circ$}\hspace{0.3em}\mbox{PM 100 pF\,$\ge$\,60$^\circ$}\hspace{0.3em}\mbox{PM 160 pF\,$\ge$\,50$^\circ$}}\hspace{0.3em}\mbox{$P$\,$\le$\,1.2\,mW}\hspace{0.3em}\mbox{$C_t$\,$\le$\,0.5\,pF}\hspace{0.3em}\mbox{SR avg.\,$\ge$\,10\,V/\textmu s}\hspace{0.3em}\mbox{SR min.\,$\ge$\,5\,V/\textmu s}\hspace{0.3em}\mbox{$T_s$\,$\le$\,14\,ns}\hspace{0.3em}\mbox{FOM$_S$\,$\ge$\,10\,k}\hspace{0.3em}\mbox{LC-FOM$_S$\,$\ge$\,20\,k}\hspace{0.3em}\mbox{FOM$_L$\,$\ge$\,0.833\,k}\hspace{0.3em}\mbox{LC-FOM$_L$\,$\ge$\,1.67\,k}\hspace{0.3em}\mbox{CMRR\,$\ge$\,60\,dB}\hspace{0.3em}{\mbox{PSRR$^{+}$\,$\ge$\,60\,dB}\hspace{0.3em}\mbox{PSRR$^{-}$\,$\ge$\,60\,dB}}\hspace{0.3em}\mbox{$|V_{os}|$\,$\le$\,1\,mV}\hspace{0.3em}\mbox{Swing\,$\ge$\,0.3\,V$_{pp}$}\hspace{0.3em}\mbox{(at swing points $|V_{GS}|-|V_{th}|\ge0$, $|V_{DS}|-|V_{DSAT}|\ge\SI{10}{mV}$)} \\
All & closed-loop stability, no right-half-plane pole at any load \\
\bottomrule
\multicolumn{2}{@{}p{\textwidth}@{}}{\rule{0pt}{2.2ex}Device rules at the operating point: $|V_{GS}|-|V_{th}|\ge\SI{10}{mV}$, $|V_{DS}|-|V_{DSAT}|\ge\SI{50}{mV}$ (except two triode devices in B2).}
\end{tabular*}

\end{table*}
\textbf{Request.} Each request contains an HSPICE netlist with
\texttt{.param} design variables, specifications, device rules, other constraints, and a measurement definition for each specification. The deliverables are a \texttt{.param} list, the operating point and
a pass/fail claim.

\textbf{Agent.} In each run, an autonomous agent receives the request and the DeX manual, and accesses DeX only through
its client. Its prompt tells it to read the manual first, continue until every spec is verified on DeX with margin or there is evidence that a spec is infeasible, and return a pass/fail claim and a summary. The agent plans the bias, assigning each device a role and a branch current, and then sizes the devices iteratively. For instance, it chooses $L$, $I_D$ and $V_{ov}$ for each device, looks up $W$, solves the circuit, measures the specs and
checks the rules, then diagnoses the violations and revises the design
(Fig.~\ref{fig:flow}).

\textbf{Judgment.} An evaluator independent of the agent re-measures the specified metrics and checks all device rules in DeX using the delivered \texttt{.param} list and operating point. The final design is then evaluated in HSPICE
(Section~\ref{sec:outcome}). A design is considered passing only if it satisfies all specifications and device rules
\cite{murmann2026isscc}.

The agent must produce a final design and report whether it satisfies all specifications and device rules. Independent evaluation checks this claim rather than relying on the model's assessment of its own work
\cite{huang2024selfcorrect}. Section~\ref{sec:results} evaluates CRAFT on four circuits using three models and examines the agents' design processes.

\section{Experiments}\label{sec:results}
\subsection{Benchmark and setup}\label{sec:setup}
Fig.~\ref{fig:schematics} shows the four benchmarks \cite{insight2025,guo2011dacfc}, designed in a commercial \SI{28}{nm}
process, with 8 to 37 transistors and 15 to 53 design variables.
\begin{figure}[t]
\centering
\begin{tikzpicture}[font=\small]
\def\reqW{\dimexpr\columnwidth-12pt\relax}           %
\def\reqC{\dimexpr\reqW-17.5mm-6.4pt\relax}          %
\node[anchor=north west, inner sep=0pt] (ttl) at (0,0) {\textbf{Request for Benchmark 4}};
\def\reqTab{\renewcommand{\arraystretch}{1.0}\setlength{\tabcolsep}{2pt}}%
\node[anchor=north west, inner xsep=1pt, inner ysep=2pt, text width=\reqW] (l1)
  at ([yshift=-4pt]ttl.south west) {\reqTab%
\begin{tabular}{@{}>{\raggedright\bfseries}p{17.5mm}@{\hspace{3pt}}!{\color{black!35}\vrule width 0.4pt\hspace{3pt}}@{}>{\raggedright\arraybackslash}p{\reqC}@{}}
Spec & $A_0\ge\SI{60}{dB}$, GBW $\ge\SI{80}{MHz}$, CMRR, \dots\\[1pt]
Device rules & $V_{ov}\ge\SI{10}{mV}$, $|V_{DS}|\ge|V_{DSAT}|+\SI{50}{mV}$\\[1pt]
Spec definition & $A_0$ is the loop gain at \SI{1}{Hz}, opened at the inverting input of the unity-gain
buffer, \dots\\[1pt]
Deliverables & \texttt{.param} list, operating point, claims
\end{tabular}};
\node[anchor=north west, inner xsep=1pt, inner ysep=2pt, text width=\reqW] (l2)
  at ([yshift=-3pt]l1.south west) {\reqTab%
\begin{tabular}{@{}>{\raggedright\bfseries}p{17.5mm}@{\hspace{3pt}}!{\color{black!35}\vrule width 0.4pt\hspace{3pt}}@{}>{\raggedright\arraybackslash}p{\reqC}@{}}
Circuit description & which device does what: PMOS-input folded cascode, class-AB output stage, feed-forward pair,
indirect compensation through a mirror, the bias stacks, the polarity, \dots\\[1pt]
Design method & separate hard constraints from performance, budget the voltages first, write the spec equations,
size with \gmid{}, measure and calibrate, \dots\\[1pt]
Hand analysis & GBW $=G_{m1}/(2\pi\,k_c C_2)$,\newline PM $=90^\circ-\sum_i\arctan(\mathrm{GBW}/p_i)$,\newline
$A_0=R_{out}\,(g_{m19}+\dots)\,G_{m1}R_{91}+\dots$
\end{tabular}};
\begin{scope}[on background layer]
\node[fill=yellow!22, draw=orange!55, line width=0.4pt, rounded corners=1.5pt, inner xsep=2pt, inner ysep=0pt,
      fit=(l2)] (l2box) {};
\end{scope}
\node[anchor=south east, fill=orange!55, text=white, font=\scriptsize\bfseries, inner xsep=2pt, inner ysep=1pt,
      rounded corners=1pt] at ([yshift=-0.2pt]l2box.north east) {L2 only};
\node[draw=black!60, rounded corners=2pt, line width=0.5pt, inner xsep=4.3pt, inner ysep=5.4pt, fit=(ttl)(l1)(l2)] {};
\end{tikzpicture}%
\caption{The Benchmark~4 request (L1). L2 adds the yellow rows.}
\label{fig:request}
\end{figure}
Table~\ref{tab:specs} lists the evaluated metrics, their limits, and the device rules. Amplifiers are biased in a unity-gain
configuration at $V_{CM}$, and their loop gain is measured with the loop opened at the inverting input. For Benchmark~3,
each metric is evaluated at all three specified loads, and its worst-case value is used. During sizing, metrics usually measured in transient simulation are estimated from DC and small-signal analysis. For example, slew rate is the current available to charge a capacitor divided by its
capacitance, read from DC solutions.
Settling and rise times come from the step response reconstructed from the closed-loop poles and zeros: the settling time is the time after which
the response stays within 1\,\% of its final value, and the rise time is its 10--90\,\% time. The HSPICE verification measures them in a transient simulation.
The device rules keep every transistor in saturation with margin, \SI{10}{mV} of overdrive and \SI{50}{mV} of
$|V_{DS}|-|V_{DSAT}|$, except two triode devices of Benchmark~2.
\begin{figure}[t]
\centering
\includegraphics[width=\columnwidth]{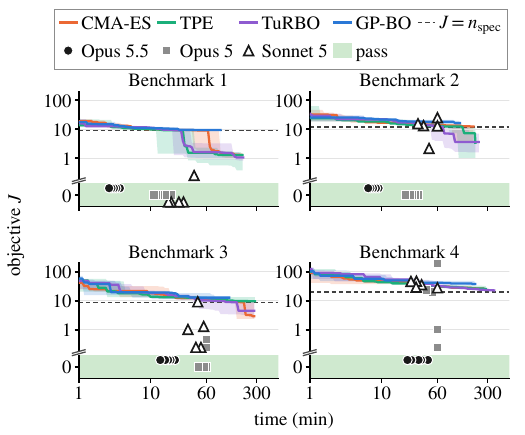}
\caption{Objective $J$ of CRAFT's final designs and the convergence curves of four optimizers (median and range
of five seeds). The agents' time-out is 60\,min.}
\label{fig:optctl}
\end{figure}
\begin{figure}[t]
\centering
\includegraphics[width=\columnwidth]{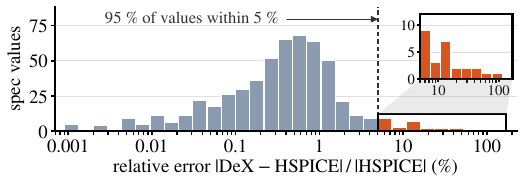}
\caption{Relative error of DeX against HSPICE on every spec value of the final designs, except $|V_{os}|$ and $|\Delta V_{out}|$, whose values are near zero.}
\label{fig:agree}
\end{figure}%
\begin{table}[t]
\caption{Hard split of AMS-SizingBench \cite{autosizer2026}.}
\label{tab:autosizer}
\centering\fontsize{7.5}{8.6}\selectfont\setlength{\tabcolsep}{2.5pt}%
\begin{tabular}{@{}lcccc@{}}
\toprule
Method & FOM$\uparrow$ & Evals$\downarrow$ & Time (s)$\downarrow$ & SR$\uparrow$\\
\midrule
Genetic & $3.4$\,{\fontsize{6.5}{7.5}\selectfont$\pm1.3$} & $165.6$\,{\fontsize{6.5}{7.5}\selectfont$\pm49.2$} & $1689.2$\,{\fontsize{6.5}{7.5}\selectfont$\pm65.1$} & 50.0\,\%\\
BO & $3.5$\,{\fontsize{6.5}{7.5}\selectfont$\pm0.2$} & $151.3$\,{\fontsize{6.5}{7.5}\selectfont$\pm51.2$} & $1875.1$\,{\fontsize{6.5}{7.5}\selectfont$\pm73.5$} & 50.0\,\%\\
TuRBO & $2.7$\,{\fontsize{6.5}{7.5}\selectfont$\pm1.9$} & $107.2$\,{\fontsize{6.5}{7.5}\selectfont$\pm19.1$} & $2105.2$\,{\fontsize{6.5}{7.5}\selectfont$\pm42.7$} & 25.0\,\%\\
\midrule
ADO-LLM~\cite{adollm2024}$^{\ddagger}$ & $3.5$\,{\fontsize{6.5}{7.5}\selectfont$\pm0.7$} & $162.3$\,{\fontsize{6.5}{7.5}\selectfont$\pm42.3$} & $1478.5$\,{\fontsize{6.5}{7.5}\selectfont$\pm68.5$} & 41.0\,\%\\
LEDRO~\cite{ledro2025}$^{\ddagger}$ & $3.8$\,{\fontsize{6.5}{7.5}\selectfont$\pm0.6$} & $140.8$\,{\fontsize{6.5}{7.5}\selectfont$\pm23.2$} & $1350.8$\,{\fontsize{6.5}{7.5}\selectfont$\pm62.5$} & 80.0\,\%\\
EE-Sizer~\cite{eesizer2025}$^{\ddagger}$ & $3.7$\,{\fontsize{6.5}{7.5}\selectfont$\pm0.9$} & $170.8$\,{\fontsize{6.5}{7.5}\selectfont$\pm20.5$} & $1872.6$\,{\fontsize{6.5}{7.5}\selectfont$\pm54.6$} & 56.4\,\%\\
AutoSizer~\cite{autosizer2026}$^{\ddagger}$ & $4.5$\,{\fontsize{6.5}{7.5}\selectfont$\pm2.7$} & $63.6$\,{\fontsize{6.5}{7.5}\selectfont$\pm62.5$} & $1473.2$\,{\fontsize{6.5}{7.5}\selectfont$\pm39.3$} & 100\,\%\\
AutoSizer~\cite{autosizer2026}$^{\dagger}$ & $6.7$\,{\fontsize{6.5}{7.5}\selectfont$\pm6.2$} & $68.4$\,{\fontsize{6.5}{7.5}\selectfont$\pm84.2$} & $521.7$\,{\fontsize{6.5}{7.5}\selectfont$\pm636.0$} & 92.3\,\%\\
\midrule
CRAFT$^{\dagger}$ & $6.4$\,{\fontsize{6.5}{7.5}\selectfont$\pm6.2$} & $\mathbf{2.9}$\,{\fontsize{6.5}{7.5}\selectfont$\pm4.6$} & $\mathbf{142.3}$\,{\fontsize{6.5}{7.5}\selectfont$\pm103.9$} & 100\,\%\\
\bottomrule
\end{tabular}

\vspace{1pt}
\raggedright\fontsize{7.5}{8.6}\selectfont $^{\dagger}$Claude Opus~5.5, three runs per circuit. $^{\ddagger}$Gemini~2.5 Flash. Rows without~$^{\dagger}$ are as reported in \cite{autosizer2026}. Values are mean~$\pm$~standard deviation. Evals: number of ngspice simulations.
\vspace{-6pt}%
\end{table}%
\begin{figure}[t]
\centering
\includegraphics[width=\columnwidth]{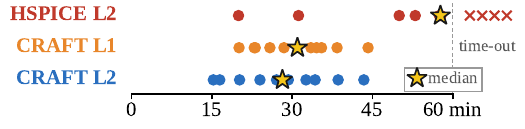}
\caption{Time from the first design step to the first passing design on Benchmark~4, ten runs each, with L1 and L2 as in
Fig.~\ref{fig:request} (medians 57.7, 31.0 and \SI{28.2}{min}).}
\label{fig:runtime}
\vspace{-6pt}%
\end{figure}%
\begin{figure*}[t]
\centering
\includegraphics[width=\textwidth]{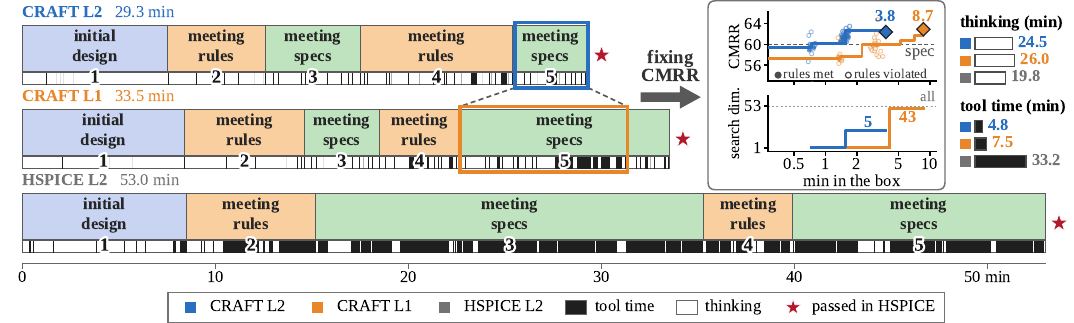}
\caption{One run each on Benchmark~4, from
the first design step to the final design: CRAFT from L2, CRAFT from L1, and the same agent using HSPICE from L2. Upper right: how the two CRAFT runs fixed the CMRR.}
\label{fig:case}
\end{figure*}%
Three Claude models (Opus~5.5, Opus~5 and Sonnet~5) at medium reasoning effort ran five times on each
of the four circuits, 60 runs in all. Each run stopped when the agent returned its design, when it reached its first passing design by its own measurements, or at \SI{60}{min}. Every run received the L1 request for its circuit, the white rows of Fig.~\ref{fig:request}: the specification, the device rules, the
measurement definition of each specification and the deliverables.

\subsection{Outcome}\label{sec:outcome}
HSPICE evaluated every final design (TT, \SI{25}{\celsius}). How far a design
$\mathbf z$ falls short of the device rules and the specs is measured by
\begin{equation}
J(\mathbf z)=\begin{cases} n_{\mathrm{spec}}+R(\mathbf z), & R(\mathbf z)>0,\\ S(\mathbf z), & R(\mathbf z)=0,\end{cases}
\label{eq:J}
\end{equation}%
where $R=\sum_j\rho_j$ sums over the $n_{\mathrm{rule}}$ device rules $u_j\ge\theta_j$ of Table~\ref{tab:specs}. Here $u_j$ is the
overdrive or $|V_{DS}|-|V_{DSAT}|$ of a device at the operating point (at each load in Benchmark~3), and its shortfall is
$\rho_j=\min(10,(\theta_j-u_j)_+/\SI{100}{mV})$. Likewise, $S=\sum_i v_i/(1+v_i)$ sums over the $n_{\mathrm{spec}}$ specs of
Table~\ref{tab:specs}, stability included, where $v_i=(\ell_i-y_i)_+/|\ell_i|$ is the relative shortfall of the HSPICE value $y_i$ from its
lower limit $\ell_i$ ($y_i-\ell_i$ for an upper limit), and the term is 1 for an unstable circuit or a value that cannot be computed.
As each term of $S$ is at most 1, $J\le n_{\mathrm{spec}}$ exactly when every
rule holds. A design that satisfies every rule therefore ranks before any that does not (the feasibility-first order of
\cite{deb2000constraint}), and a value of $J=0$ indicates a passing design. Fig.~\ref{fig:optctl} shows $J$ of the agents' final designs and of four black-box optimizers on every
benchmark. Opus~5.5 passed in all 20 runs, Opus~5 in 13 and Sonnet~5 in 4, so the easier the benchmark and the stronger
the model, the more runs passed. The optimizers, CMA-ES \cite{hansen2001cmaes}, TPE \cite{bergstra2011tpe}, TuRBO \cite{eriksson2019turbo} and Gaussian-process Bayesian optimization \cite{snoek2012bo}, ran on HSPICE over the
same design variables, each minimizing $J$ of \eqref{eq:J} from no initial sizing, with five seeds per benchmark and
1000 candidates per run (300 for the Gaussian process). No optimizer run produced a design satisfying all specifications. On Benchmark~4, no optimizer run satisfied all device rules. As a positive control, CMA-ES and TuRBO, started from a passing design
with only its bias moved and run with a narrower step, reached $J=0$ within 100 candidates in 18 of 32 runs (four starts for each optimizer on each benchmark), and in at least 3 of 8 on every
benchmark.

Excluding $|V_{os}|$ and $|\Delta V_{out}|$, which are near zero, 95\,\% of the metric values of all
final designs lie within 5\,\% of the HSPICE values (Fig.~\ref{fig:agree}). For these two, whose limits are 1 and \SI{7}{mV}, DeX is within 0.1 and \SI{1.7}{mV} of HSPICE, respectively. The remaining 5\,\% are values of the line regulation, the swing, the rejection ratios
and a few specs of failing designs. Because the line-regulation values are near zero, small absolute differences produce large relative errors. The swing is the output range over which every device satisfies its rule, so its boundary lies where some device
reaches its rule limit. When that device's margin changes slowly with the output, a small margin difference is amplified into
a large swing difference. Apart from line regulation and swing, the relative errors among these outliers are within 20\,\%. None of the discrepancies exceeding 5\,\% changed a design's overall pass/fail classification.

\subsection{Comparison with baselines}\label{sec:optctl}
Table~\ref{tab:autosizer} compares CRAFT with three optimizers and four LLM sizing methods on the Hard split
of AMS-SizingBench, following the setup of AutoSizer \cite{autosizer2026}. The circuits are designed in SKY130, and
DeX uses SKY130 tables for them. The request gives the specification and asks only that it be met. Here SR is the success rate in ngspice, and FOM and Time are defined as in AutoSizer. The agent used ngspice only for some specs, such as those from a
transient analysis, and Evals counts these ngspice runs and the final ngspice check. CRAFT met the specification in every run, with far fewer Evals and in about a quarter of the time taken by AutoSizer$^{\dagger}$ on the same LLM.

\subsection{How the agents designed}\label{sec:behaviour}
Fig.~\ref{fig:runtime} shows thirty runs by Opus~5.5 on Benchmark~4, the largest circuit: ten
with CRAFT from L1, ten with CRAFT from L2, and ten with HSPICE in place of DeX from L2. L2 adds the yellow rows of
Fig.~\ref{fig:request}. Each run stopped as in Section~\ref{sec:setup}.
With CRAFT from L2, every run passed, with a median of \SI{28.2}{min}. With HSPICE, six of ten passed, with a median of \SI{57.7}{min} including the four time-outs. The sample median is the average of the fifth and sixth ordered times, both observed before the \SI{60}{min} cutoff. CRAFT thus cut the median time to a passing design by 51\,\%.
From L1, CRAFT also passed in every run, with a median of \SI{31.0}{min}.

Fig.~\ref{fig:case} follows one run each of CRAFT from L2, CRAFT from L1 and the HSPICE agent from L2, which reached the final
design in 29.3, 33.5 and \SI{53.0}{min}, respectively. All three made an initial design and then alternated between resolving device-rule
violations and meeting specifications, returning to whichever was violated again. For example, the run of CRAFT from L2 proceeded in five stages. In stage~1 it budgeted the headroom, choosing each length so that every device
would fit under the supply, planned the node voltages and the branch currents from the device rules and the power budget,
and read the widths from the tables. In stage~2 it reduced the rule violations and deferred the rest; in
stage~3 it fixed the phase margin by adjusting the compensation at the limiting node; in stage~4 it cleared the rule
violations with an agent-generated parameter-tuning script (tuner) to adjust the sizes and bias; and in stage~5 it fixed the CMRR by resizing the tail
devices. The two CRAFT runs differed most in fixing the CMRR, which took the L2 run
\SI{3.8}{min} and the L1 run \SI{8.7}{min} (Fig.~\ref{fig:case}, upper right). This difference may reflect the additional CMRR guidance in the L2 request. L2 explains what limits the CMRR of this circuit: a common-mode input slightly changes the tail current, and a few paths carry this change to the output. This guidance focuses the search on a few devices. The L2 run ranked the devices by their effect on the CMRR within \SI{0.5}{min}, tried single-variable changes and then searched only 5 of the 53 variables together. The L1 request only defines how the CMRR is measured; the L1 run changed one variable at a time, found the tail devices to be the cause after \SI{4}{min}, and then searched 43 of the 53 variables with a tuner. The HSPICE agent followed a similar logic with a small table it built from the simulator, and violated the
phase-margin and output-swing constraints when its tuner resolved the rule violations (stage~4). It ran about 900 simulations. Both CRAFT runs completed sooner than the HSPICE run, without SPICE simulations in their sizing loops.

\section{Conclusion}
CRAFT lets an LLM agent design an analog circuit the way a designer does, from a specification and device rules, by calculation from pre-computed device tables. Final designs are evaluated independently of
the agent. DeX answers the device queries of the lookup-table method and runs a Newton solver for the whole
netlist through one text interface, so that the specs from DC and small-signal analysis need no simulator during the design. Across four \SI{28}{nm} circuits, including
a 37-transistor three-stage amplifier with 53 design variables, all 20 Opus~5.5 runs produced designs that met every specification and device rule in HSPICE. On the largest circuit, the same agent with DeX in place of HSPICE reached a passing design in every
run, with a median time half that of the agent using HSPICE.

\bibliographystyle{IEEEtran}
\bibliography{refs}

\end{document}